# Water-resistant carbon nanotube based strain sensor for monitoring structural integrity

Preety Ahuja[1], Shingo Akiyama[1], Sanjeev Kumar Ujjain[1], Radovan Kukobat[1], Fernando Vallejos-Burgos[1], Ryusuke Futamura[1], Takuya Hayashi[2], Mutsumi Kimura[3], David Tomanek[4], Katsumi Kaneko[1,*]

[1]Center for Energy and Environmental Science, Shinshu University, Nagano, Japan

[2]Department of Water Environment and Civil Engineering, Faculty of Engineering, Shinshu University, Nagano, Japan

[3]Faculty of Textile Science and Technology, Shinshu University, Ueda, Japan

[4]Physics and Astronomy Department, Michigan State University, East Lansing, Michigan 48824, USA

Monitoring structural integrity during and after extreme events such as an earthquake or a tsunami is a mundane yet important task that still awaits a workable solution. Currently available stress sensors are not sufficiently robust and are affected by humidity. Insufficient information about crack formation preceding structural failure increases risk during rescue operations significantly. Designing durable stress sensors that are not affected by harsh and changing environment and do not fail under catastrophic conditions is a fundamental challenge. To address this problem, we developed a stress sensor based on creased single-walled carbon nanotubes (SWCNTs) encapsulated in a non-fluorinated superhydrophobic coating. The creased SWCNT film was fabricated and integrated in polydimethylsiloxane (PDMS) to provide a highly linear response under elastic deformation. The non-fluorinated water-repellent coating was fabricated by spray-coating the film with nanosilica particles, providing water resistance during elastic deformation. The compact design and superior water resistance of the sensor, along with its appealing linearity and large stretchability, demonstrates the scalability of this approach for fabricating efficient strain sensors for applications in infrastructure and robotic safety management as well as advanced wearable sensors.

Monitoring structural integrity during and after extreme events such as an earthquake or a tsunami is a mundane yet important task that still awaits a workable solution. Additionally, the mechanical frame strength of transportation must be continuously monitored for sufficient safety. Currently available strain sensors are not sufficiently robust and are affected by humidity. A water-proof strain sensor would be applicable for infrastructure safety management in harsh environments. Such a harmless water-proof strain sensor could also be used as an advanced wearable sensor. Insufficient information about crack formation preceding structural failure increases risk during rescue operations significantly. To address this problem, we developed a strain sensor based on creased single-walled carbon nanotubes (SWCNTs) encapsulated in a non-fluorinated superhydrophobic coating. The SWCNT film was fabricated and integrated in polydimethylsiloxane (PDMS) to provide a highly linear response under elastic deformation. The non-fluorinated water-repellent coating was fabricated by spray-coating the film with nanosilica particles, providing water resistance during elastic deformation. The compact design and superior water resistance of the sensor, along with its appealing linearity and large stretchability, demonstrates the scalability of this approach for fabricating efficient strain sensors for applications in infrastructure and robotic safety management.

Governments must guarantee safety and present environments to preserve life while maintaining a comfortable urban landscape with minimal economic burden. Particularly, numerous infrastructures, such as buildings, bridges, dams, and tunnels, in advanced countries have serious deterioration issues. Additionally, the mechanical frame strength of transportation means, such as cars, aircrafts, ships, and trains, must be continuously monitored for safety. Consequently, an internet-mediated sensing system for detecting the sudden and local deterioration of infrastructures needs to be established to support appropriate renovations; such a system would require the use of strain sensors that are highly stable even under harsh environments. The development of high-sensitivity water-proof strain sensors is urgently needed; in particular, non-fluorinated water-proof sensors are preferable for assuring safety. Currently, there are no active studies on sensors for monitoring structural integrity, although there have been many reports on wearable sensors[1,2], which have wide applications in the medical field (e.g., robotic skin and health monitoring), the military, and even in the film and game industries for capturing human motions. Some of these sensors can be efficiently applied for monitoring structural integrity with enhanced environmental tolerance[3].

However, the sensing ability of existing sensors is limited in terms of underwater detection, and only a few reports have suggested sensors with underwater sensing capability[4–6]. A superhydrophobic $WS_2$ nanosheet-wrapped sponge has been reportedly fabricated by dipping a sponge in commercial hydrophobic fumed silica nanoparticles[5]. Jeon et al. encapsulated their sensor (SWCNT/Ag composite with PDMS) in a flexible fluoropolymer and presented its sensitivity to bending in water[4]. Multiwalled carbon nanotubes composite (MWCNTs/$Fe_3O_4$) on thermoplastic elastomer (TPE) have been used as a superhydrophobic coating with a strain tolerance of ~50%[7]. Among these, in situ approach for producing water resistive sensors are still associated with limitations involving applications in harsh environments. Hence, the simple coating process for fabricating superhydrophobic sensor is promising. However, chemically safe superhydrophobic smart coatings for stretchable sensors with sustained piezoresistive responses has yet to be realized.

Carbon nanotubes represent the foundation for most of the conventional stretchable strain sensors[1,8–13], owing to performance characteristics which considerably exceed from known metal/metal oxides or organic semiconductors based sensors[14–16], including large stretchability, high piezoresistivity and long term stability under mechanical strain. More precisely, single-walled carbon nanotubes (SWCNTs) possess one dimensional morphology with high aspect ratio to offer high internal piezoresistivity for use in strain sensors[17,18]. The main challenge in its use as sensing element follows from the restrictions associated with the fabrication strategies of the sensor[19–25]. Fragmented SWCNT paper with cracks has been integrated into PDMS, which endows the device with a large gauge factor of ~$10^7$ in the sensing region up to 50%[26]. Thickness gradient SWCNT film has been reported and exhibited segmented sensitivity with substantial structural changes under applied strain , thereby persuading non-linear response[27]. Wrinkled CNT film has been reported with coffle(highly stretchable elastomer) but still suffers from sensitivity deviations[28]. These approaches yield sensors with very high sensitivity at large strains leading to exponential increment in piezoresistive response. In addition, an enhanced initial resistance and a non-linear response (sensitivity deviation) are always associated, increasing the need of the complicated calibration[28]. This may attributed to the release of strain energy via fracture of CNT junctions ensuing unsteady response[29–32]. The difficulty in fabrication of SWCNT based sensors of linear response emanated from ill contact between SWCNT owing to insufficient dispersion of SWCNT bundles in PDMS. We must disperse SWCNT uniformly in PDMS media to produce stretchable SWCNT networks with good electrical contact. Newly developed Zn-Al sol gel dispersant for SWCNT bundles developed by our group is promising to embed stably the network of SWCNT in PDMS[33].

Here, we report the uniform and fine dispersion of SWCNT in PDMS, enable to fabricate stably stretchable SWCNT networks to provide the strain sensor of excellent linear response. The good electrical contact between SWCNT is guaranteed by molecular level anchoring of SWCNT in PDMS, stem from weak charge transfer interaction as demonstrated by in-situ Raman spectra. To date, Raman spectra are utilized for the first time to gain in-depth understanding of the concerted motion of SWCNT with PDMS in response to applied strain to detect the microscopic changes very sensitively. Resulted in-situ Raman spectra in correspondence with the optical micrographs provide the evidence regarding stabilized interaction between SWCNT and PDMS, leading to linear piezoresistive response. Moreover, the continuous usage of the fabricated sensor augmented the performance of the sensor owing to stabilization process, steered to SWCNT stacks along with reproducible and reliable response. This sensor was further coated with chemically safe nanosilica to confer marked water resistance and is thus a powerful sensor for monitoring structural integrity under harsh environments.

**Water-proof Strain Sensor Fabrication**. Firstly, SWCNTs were dispersed in water with a Zn-Al sol-gel dispersant following a previously described protocol[33] and utilized for making water resistive PDMS embedded strain sensor (PDMS/SWCNT/PDMS) as shown in Supplementary Fig. S1.

Newly developed sensor exhibits contact angle of 161.5°±0.8 and this superhydrophobicity is remarkably maintained even when stretched with 100% strain as shown in Fig. 1a. The effectiveness of the superhydrophobic coating is evidented by lesser contact angle (108°±2) of sensor without the water-repellent coating. It is further demonstrated in Supporting Video V1 by its ability to block the water jet falling on the sensor in both, the relaxed and stretched states (Supplementary Fig. S2). No water residue remained on the surface of the water-proof sensor after testing under the water jet in contrast to the bare sensor (without water-repellent coating), which guarantees the robust water-repellent coating of the sensor. Notably, the water-repellence was maintained in corrosive environment, such as in acidic, alkaline or saline aqueous solutions (Supporting Video V1) evidencing their applicability in harsh environments. We also trace the stability of the superhydrophobic coating on the sensor by storing the sensor in water for 100 h. We found that the contact angle did not change even after 100 h, suggesting the reliable applicability of the sensor under water (Fig. 1b). In addition, the piezoresistive response of the water-repellent sensor was observed under humid conditions by packing the sensor between wet sponges (Fig. 1c) and subjecting it to consecutive loading-unloading cycles to100% strain. The arrangement of the sensor between dry and wet sponges is illustrated in Supplementary Fig. S2. The response to the applied strain showed no piezoresistive change (within reproducibility) under the dry and wet conditions over large stretchability range (Fig. 1d), demonstrating successful deposition of the superhydrophobic coating on the fabricated sensor. The actual response of the sensor with and without the superhydrophobic coating under humid condition is shown in Supplementary Fig. S2, endowing the efficient water resistive strain sensor. Such a performance has not been observed previously in CNT based stretchable strain sensors.

**Comparative piezoresistive performance.** Besides superhydrophobicity, the realization of highly linear piezoresistive response may now be possible. Previously, others have reported highly stretchable CNT based strain sensors but they are associated with large sensitivity deviations. For further understanding, the comparative piezoresistive performance of the fabricated sensors published in the literature on SWCNT based strain sensors are demonstrated in Fig. 1e and 1f. Here, the plots begin above 3 % strain due to lack of data in literature and are made to fall in scale to show the actual response. The sensor reported by Z. Liu et al[27] gives linear response at lower strain with high relative resistance change as shown in Fig. 1e. However, the linear plot deviates in the large strain range. The response curve by Park et al.[28] shows extremely low relative resistance change. The residual one by Zhou et al.[26] shows high relative resistance change along with strain tolerance of 50% with non-linear response. Reliable and linear response are the key to reproducibility of stretchable strain sensor. Our sensor exhibits a complete linear response passing the origin with the consistent sensitivity, even below 1% strain. Also, we need to discuss the performance of the sensor with regard to sensitivity (gauge factor). Our PDMS/SWCNT/PDMS sensor shows almost no change in gauge factor over the whole stretchability range (Fig. 1f), ascertaining the linear response of the sensor. However, the gauge factor reported in literatures[26,27] showed large deviation with the strain. The gauge factor in the literature[28] is almost constant but the absolute value is too small. Consequently, the present PDMS/SWCNT/PDMS has the excellent performance in comparison with the previously reported sensors.

**Concerted interaction between SWCNTs and PDMS under strain.** The effect of structural deformations on the fabricated sensor was evaluated by observation with in situ optical microscopy (Fig. 2a-2f). The concerted motion of the SWCNTs with the PDMS in response to applied strain demonstrates the opening of the creases at lower strains, along with crack formation and propagation at higher strains. It is evidently shown that the well dispersed SWCNT in PDMS enabled stretchable connection throughout the cracks, leading to good electrical contact, even at higher strains. In addition, these cracks vary linearly with the applied strain (Fig. 2g) while maintaining stable electrical contact during strain loading-unloading cycle, causing a linear sensor response. The first loading-unloading cycle stabilizes the inter SWCNT contacts in PDMS, thereby inducing constricted cracks (Supplementary Fig. S3 a-e) giving initial irreversible response (Fig. 2g). The second loading-unloading cycle produce no new cracks (Supplementary Fig. S3 f-j), leading to the reversible and reproducible linear response.

The stretching decreases the electrical contact area between the well-connected SWCNT resulting in gentle resistance increments without discontinuation of the electrical conducting networks at higher strains, thus allowing a highly linear piezoresistive response over a wide stretching range. Furthermore, the effect of continuous stretching and releasing during several cycles was observed by optical recordings (Fig. 2d-2f) at different strain. A more stabilized network of SWCNT is formed, in which some of the SWCNTs interact more strongly with each other and undergo structural variations, resulting in more agglomerated stacks. Notably, the wide and short cracks (Fig. 2c) formed during the initial loading-unloading process at 100% strain become steadier with narrower counterparts (Fig. 2f) due to weaker PDMS intercalation. Consequently, a more stabilized response is attained after continuous loading-unloading cycles.

The parallel contact of great stability must be guaranteed by molecular level contact between SWCNT and PDMS. We recorded in-situ Raman spectra in accordance with the optical images for two areas, i.e., black and white areas resulting from crack propagation. The black and white areas stem from different SWCNT density distributions in response to the applied strain. At higher strains, well connected SWCNTs have lesser contact area with reduced density in particular regions, imparting more transparency (white area) due to PDMS intercalation. This presence of SWCNTs at white area is excitingly evidenced by the existence of G-band peaks at higher strains in the Raman spectra (Supplementary Fig. S4). Correspondingly, we measured the Raman spectra of black and white areas to obtain insight regarding the strain effect. The G-band peak of the SWCNT in the black area is observed at 1604 $cm^{-1}$ (Fig. 2h) due to the in-plane vibration of carbon-carbon bonds, whereas the G-band peak of bare SWCNTs is at 1589 $cm^{-}$ [1].This blue shift versus the pristine SWCNTs is attributed to the charge transfer from the SWCNTs to the PDMS.[33–36]

The G-band peak shift for the black and white areas versus strain is shown in Fig. 2i. The black and white areas exhibit the same G-band peak position at the initial stage up to 30% strain. The extent of the charge transfer interaction between the SWCNTs and the PDMS on strain manifests the shift in the G-band peak position. The black area shows an up shift to 50% strain followed by a downshift at higher strains, while the white area displays downshifting beyond 30% strain. This behaviour is different from the literature result indicating strain causes a simple downshift in the G-band peak position[37–39]. Our experiment clearly shows that the black and white areas play different roles. We must take into account two factors regarding the different shift behaviours of the G-band peak position with strain. One is the charge transfer interaction of the SWCNTs with the PDMS surface due to the disentanglement of SWCNTs in the film. The other is the isolation of the SWCNTs from the PDMS with weaker charge transfer interactions. In the black area at lower strains, well dispersed and entangled SWCNTs undergo disentanglement[33] by dissipating strain through creases and thus interact more strongly with the PDMS. However, the latter factor dominates at higher strains, inducing weaker charge transfer interactions between the isolated SWCNTs and PDMS. This change consequently weakens the interaction among nearby SWCNTs and lower their vibration frequency[40]. A similar G-band peak shift is observed at lower strains up to 30% for white area; the lower SWCNT content and weaker contact of the SWCNTs with the PDMS induces a greater downshift in the G band. The G band for the black and white areas shows similar slight downshifting in response to the applied strain after several repeated loading-unloading cycles. This gradual shift of the G-band peak position of SWCNT indicates the stably associated SWCNT contacts and corresponding less charge transfer interaction with PDMS. Therefore, we can obtain a good linear piezoresistive response to strain.

**Linear response of electrical resistance against strain.** The linear I-V characteristic of the embedded sensor (PDMS/SWCNT/PDMS) demonstrates ohmic contact (Supplementary Fig. S5), able to provide a stable electrical output signal[41], enabling to employ the assembled configuration as a strain sensor. The sensor was subjected to 100% strain by stretching machine at a rate of 1 mm per sec to analyse the sensor performance and the corresponding change in resistance was recorded. The sensor was firmly fixed between Pt mesh to keep stable electrical contact with 1 cm$^2$ as an effective sensor area for stretching (Supplementary Fig. S6). Fig. 3a demonstrates the state of the sensor when strained (100%) and unstrained (0%). The resistance change with strain during loading and unloading follows monotonic and linear increments following the strain path (Fig. 3b). This one-to-one correspondence in resistance output with the applied strain makes the device suitable for effectively detecting a strain[42] during infrastructures monitoring. The behaviour of the initial loading-unloading cycle is different from that of subsequent cycles owing to the irreversible resistivity loss during the first cycle, which was attributed to the leftover narrower cracks during the unloading of the first strain cycle as observed earlier. Further cycles are demonstrated in Fig. 3c, illustrating a stabilized relative resistance change with strain in the initial five strain loading-unloading cycles. The relative resistance change with strain represents the sensitivity of the sensor in terms of the gauge factor (GF)[43]. The sensitivity (GF) of the sensor at 100% strain was found to be constant, as shown in the inset Fig. 3c during the initial cycles. The linear resistance change with strain was also observed across the whole stretchability range of 100% ($R^2$ 0.97), even after five cycles of stabilization (Fig. 3d). Thus, the fabricated sensor possesses high sensitivity (GF ~2.6), not only at 100% strain but also at lower strains (Fig. 3d inset), compared to the earlier reported values[7,44,45]. The sensitivity of the sensor at every strain is similar (inset, Fig. 3d), evidencing the linear response of the fabricated sensor. The linear strain-induced resistance change through the change in the SWCNT-SWCNT contact area change with the strain, is promising for infrastructure monitoring.

An additional advantage of the present sensor (PDMS/SWCNT/PDMS) is its high sensitivity to detect lower strains with a fast response. This sensor can detect as little as 0.1% strain and as much as 100% strain with a highly linear and stable response (Fig. 3e), even using only the linear relationship between resistance and lower strain (Fig. 3e inset). The application of field effect transistors to this strain sensor should easily enable the detection of strains less than 0.01%. Furthermore, the response and recovery time are measured by loading the sensor with a quasi-transient step (5% strain at 16 mm per sec) and holding the strain for a particular time to determine the drift[46]. The corresponding resistance change during loading and unloading strain occurred within less than 100 ms (Fig. 3f), demonstrating the ultrafast response and recovery properties facilitating the detection of strain at very low angle for structural monitoring. In addition, only ~2% drift in the resistance was observed during initial stage of strain holding followed by no remarkable fluctuation, which is attributed to the viscoelasticity of PDMS[47]; this drift can be compensated using double sensors.

**Random Coil to Floc Structural Transformation of SWCNT Network.** The resistance variation mechanism of the present sensor (PDMS/SWCNT/PDMS) originates from stretchable SWCNT network with excellent electrical connection in PDMS. As the SWCNTs are highly dispersed on the PDMS in the form of fine bundles of 40 nm in the size before straining[48], the SWCNT sheet has a two-dimensionally developed random coil-like structure (Fig. 4a). The fine SWCNT coils begin to partially orient towards each other in the PDMS on applied strain; the partially oriented SWCNT coils are in contact with the surrounding PDMS through the observed charge transfer interactions. The repeated strain application induces flocculated structures consisting of the fine SWCNT bundles, as shown by Transmission Electron Microscopy (Fig. 4b). The SWCNT coils grow into a stationary slip-flow state weakly supported by the PDMS, as suggested by the smaller up shift of the Raman G-band peak, observed earlier. The SWCNT network corresponding to the flocculated structures of SWCNT and PDMS, as confirmed by TEM observation (Fig. 4b1 and 4b2, before and after cycling, respectively), provides the linear response of the electrical resistance to the strain. This electrical contact is approximated by the following simple model, which neglects the capacitive factor.

The electrical conduction path of the SWCNT/PDMS sensor during a stationary response is governed by the flocculated structures consisting of SWCNT/PDMS blocks, the block-block bridging contacts, and vacant regions, as shown in Supplementary Fig. S7. The total electrical resistance is simply described by the electrical resistance R between two arrays of the SWCNT/PDMS blocks with inherent resistance $R_I$, as given by

$$R = \frac{2R_I R_B + 2R_I R_V + R_B R_V}{R_B + R_V} \qquad (1)$$

Here, the gap resistance between the two arrays is expressed by the parallel resistance arrangement of the bridging contact resistance, $R_B$, and the resistance of the vacant region, $R_V$. The initial resistance reflects the inherent resistance, $R_I$, of the SWCNT/PDMS block. The formation of cracks in the film with strain increases the contribution of the parallel resistances of $R_B$ and $R_V$. At lower strains, the whole resistance R is governed by only $R_I$ as

$$R = 2R_I \qquad (2)$$

This is because the contribution of $R_V$ is negligible as there are insignificant vacant regions at lower strains.

Crack propagation dominates at higher strains augmenting $R_V$. As $R_V \gg R_I$, $R_B$ implies that R is governed by $R_B$.

$$R = 2R_I + R_B \qquad (3)$$

Here, $R_B$ should be proportional to the decrease in bridging contact with the increase in strain. The experimentally observed linear relationship between the resistance and strain, $\varepsilon$ (Supplementary Fig. S7), leads to

$R_B = 22.7 \times \varepsilon$ (ohm)

Consequently, $R_B$ and R at the 100% stretching are derived to be 2270 ohm and 5020 ohm respectively, whereas experimentally observed R at the 100% strain was 5125 ohm, supporting the above model.

**Linear response performance after long cycling.** The PDMS/SWCNT/PDMS sensor demonstrated stable performance, even after 15,000 cycleupto100% strain, as shown in Fig. 5a. The precise resistance change is given in Supplementary Fig. S8 and shows increment in resistance due to the stabilization process during the initial cycles, followed by steady performance. This finding was further analysed by recording the resistance change with strain after certain cycles, as shown in Fig. 5b. After stabilization (7000 cycles), the sensor maintains a high linearity with augmented response with Gauge factor ~5, especially at lower strains versus initial cycles, being indispensable to the previous sensors reported in infrastructural monitoring. Also, no further change was observed during its usage after stabilization, sustaining the highly linear response with negligible sensitivity deviations in large strain spectrum of 100%. These aforementioned characteristics imparts distinguished performance of the fabricated sensor as compared to the literature (Supplementary Table S1).

The morphological changes during the loading-unloading cycles were analysed by the SEM observation, before and after 15,000 cycles (Supplementary Fig. S9). The crease density in the PDMS/SWCNT/PDMS sensor increases from $400 \pm 17$ to $700 \pm 12$ per $mm^2$ after these cycles. The increase in the crease density is key for producing a stable response, because the SWCNTs are associated with each other in the crease, providing the stable conducting paths, as shown by SEM (inset, Supplementary Fig. S10). Additionally, the 2D profile from SEM image generated using Image J shows the increased crease density after cycling (Supplementary Fig. S10); the long cycling produces new fine creases to form stable SWCNT networks.

**Real-time Responses.** In order to determine the strength of the fabricated sensor, it is used for detecting strain in human finger motion and a stainless steel strip for real time applications in healthcare and infrastructural monitoring. We attached the prototype sensor to the forefinger of a glove via adhesive (copper tape) and used it to characterize the motion of a human finger as an example of sensing the real strain. Our sensor demonstrated a reproducible change in resistance in response to strain applied by finger motion. The bending and relaxing of the finger was sensitively detected, as shown by the increase/decrease in the resistance (Fig. 6a-6c, Supplementary Video V1). In addition, a light-emitting diode (LED) can be modulated by this sensor, thus demonstrating its potential for application in intelligent visual control systems. We connected the LED with the fabricated sensor, and the illumination intensity varied in response to the strain applied to the senor (Supplementary Fig. S11). Remarkably, the initial resistance of the sensor was low enough to light the LED using a3 V battery (Supplementary Video V1); the LED intensity sensitively depended on the strain applied. The fabricated sensor can also monitor the strain of a stainless steel strip (inset), revealing the sensitivity of the sensor to a change in angle of less than 1° (Fig. 6d), which shows its potential efficacy in the regular inspection or structural monitoring of infrastructures.

**Conclusions**

We have successfully fabricated well dispersed SWCNT film and integrated them in PDMS. Resulting SWCNT embedded strain sensor (PDMS/SWCNT/PDMS) provides stretchable SWCNT network owing to newly developed Zn-Al dispersant. This allows parallel contact between SWCNT in response to applied strain with decrement in contact area leading to linear increase in resistivity; capable of sustaining large strains (100%) with a highly linear response. This consequently prevented the failure of the electrically conductive network at higher strains, which has been associated with non-linear or exponential piezoresistive sensor response, as mostly reported in the literature. The structural transformation of the SWCNT into flocculated structures under continuous strain loading and unloading resulted in more stabilized and augmented response, especially at lower strains, evidencing their capability for efficient infrastructural monitoring. More importantly, the water resistance of such a sensor will open many new pathways for a range of applications. We used non-fluorinated chemistry involving $SiO_2$ and PDMS to generate a low-energy surface, refraining from the utilization of environmentally hazardous materials to create a water-resistant coating on the sensor, even at 100% strain. This coating allows the sensor to function efficiently under humid, acidic, saline and alkaline conditions.

**Methods**

Methods, including statement of data availability and any associated references are available in Supplementary information.

**Acknowledgements**

This research was supported by the NEDO Project for Robot Core Technology on Next Generation Artificial Intelligence.

**Author Contributions**

**Additional Information**

Supplementary information is available and available free of charge via the internet at the URL http://www.rsc.org/suppdata/c9/ta/c9ta06810d/c9ta06810d1.pdf . A supplementary movie is linked at

http://www.rsc.org/suppdata/c9/ta/c9ta06810d/c9ta06810d2.mp4 .

**Figure captions**

**Figure 1** (a) Contact angle variation at different applied strain. Inset shows an image of the contact angle down to a particular strain. (b) Time course evaluation of the contact angle on the water-repellent sensor. (c) Image of the sensor between wet sponges and (d) Resistivity change of the water-repellent sensor in response to applied strain when in contact with dry and wet sponges; Comparative performance of fabricated sensor in terms of (e) Relative resistance change in response to applied strain and (f) gauge factor with existing literature of SWCNT based stretchable strain sensors.

**Figure 2** In situ optical micrographs of sensor under different strains (a-f). Average crack opening during initial cycles at particular strains (g). Raman spectra showing variations in G-band peak position (h) and intensity (i) under different strains.

**Figure 3** (a) Image of sensor at 0% and 100% strain. (b) Relative resistance change and strain variation with time during initial cycling of sensor. (c) Resistivity change with strain for six consecutive initial cycles. (d) Resistivity and relative resistance change of sensor with strain (5$^{th}$cycle) showing a highly linear response. (e) Normalized resistance change with increasing strain. (f) Response time of the sensor during quasi-transient loading (5% strain at 16 mm sec$^{-1}$).

**Figure 4** (a)Model demonstrating stretching of the sensor to describe the deformation of the creases at lower strains and crack propagation at higher strains in agreement with TEM images (b1 and b2, before and after cycling, respectively).

**Figure 5** (a) Normalized resistance change of the sensor after 15,000 cycles in response to applied strain of 100% and (b) Relative resistance change with strain after particular cycles during cycling stability testing of the sensor.

**Figure 6** Real-time response of the sensor in detecting human finger motion (a-c). (d) Resistance change during strain detection (change in angle, degree) of a stainless-steel strip connected to a fabricated sensor.

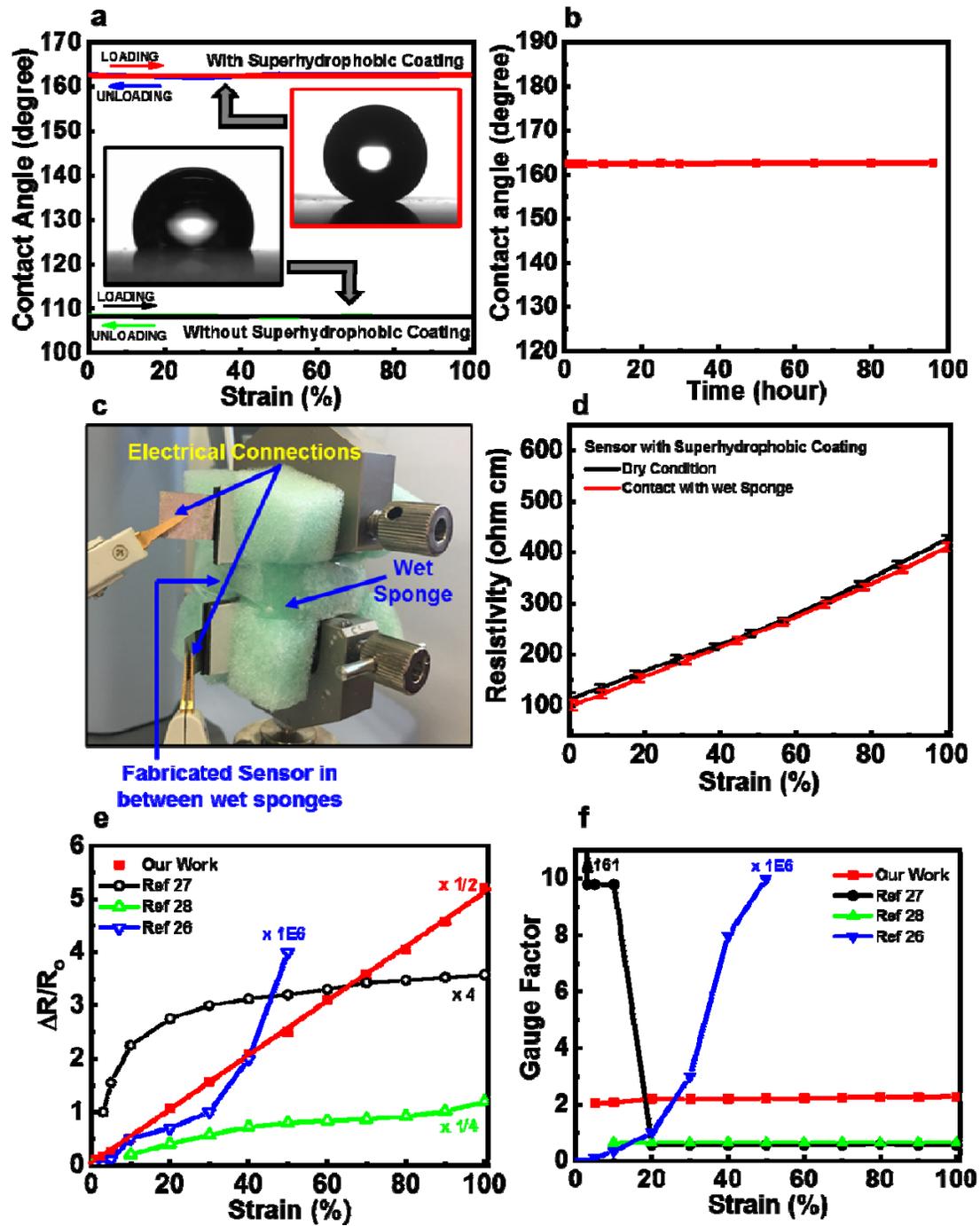

**Figure 1**

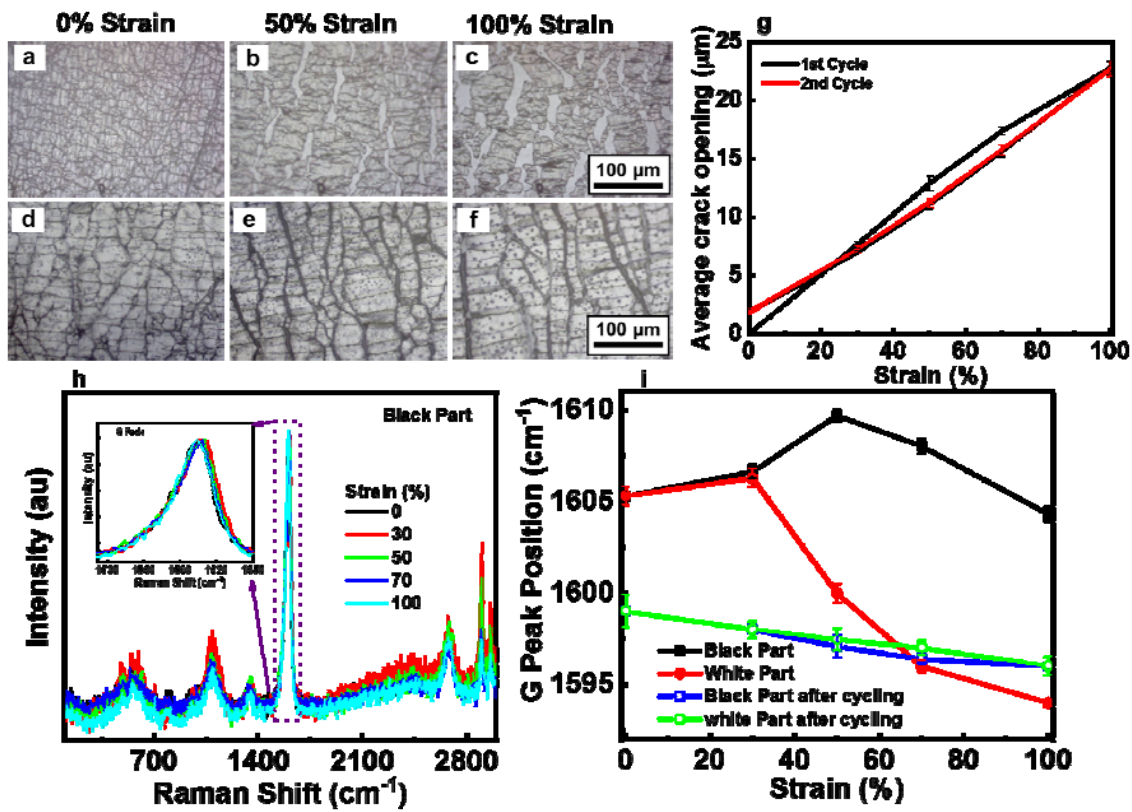

**Figure 2**

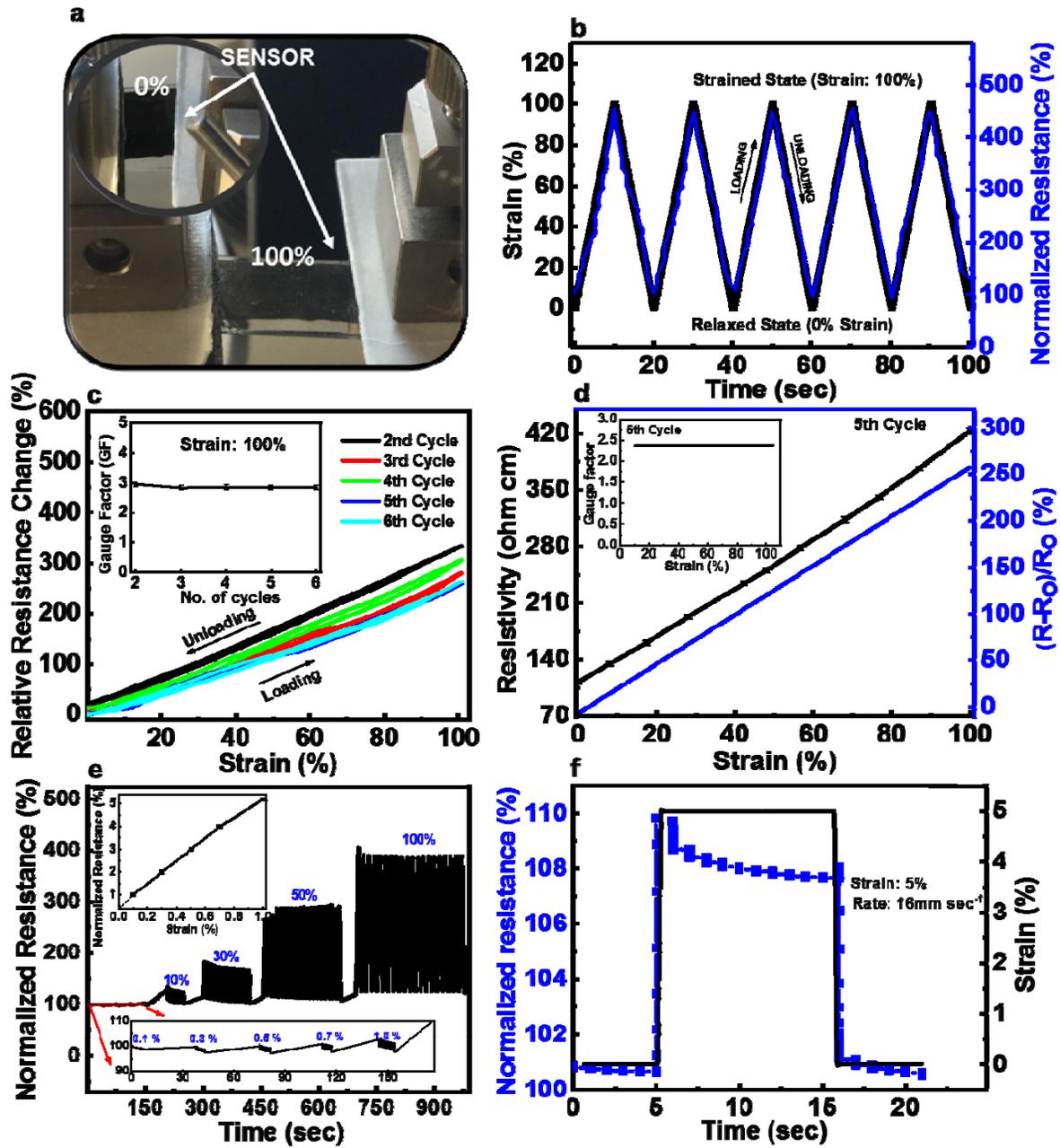

**Figure 3**

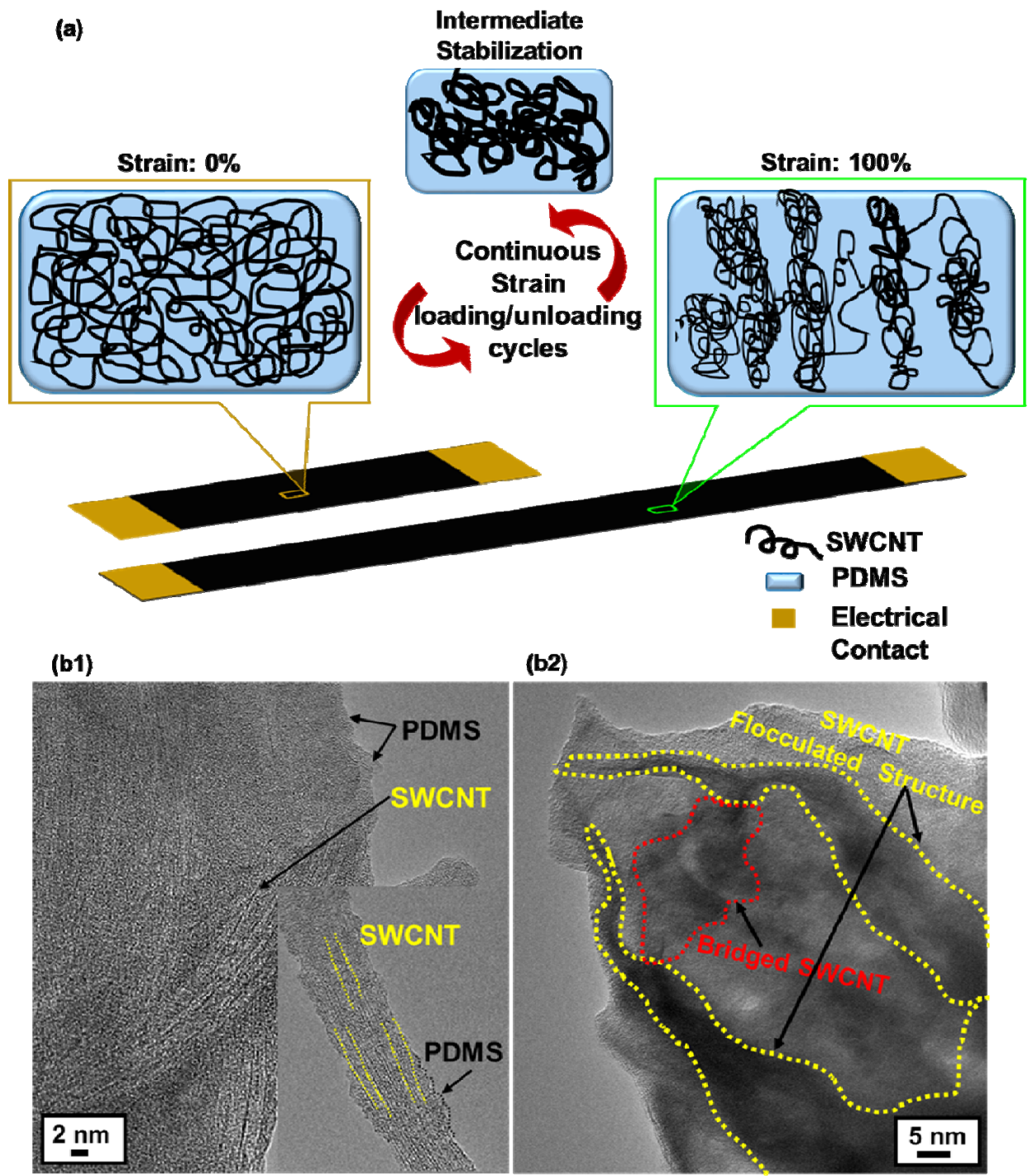

**Figure 4**

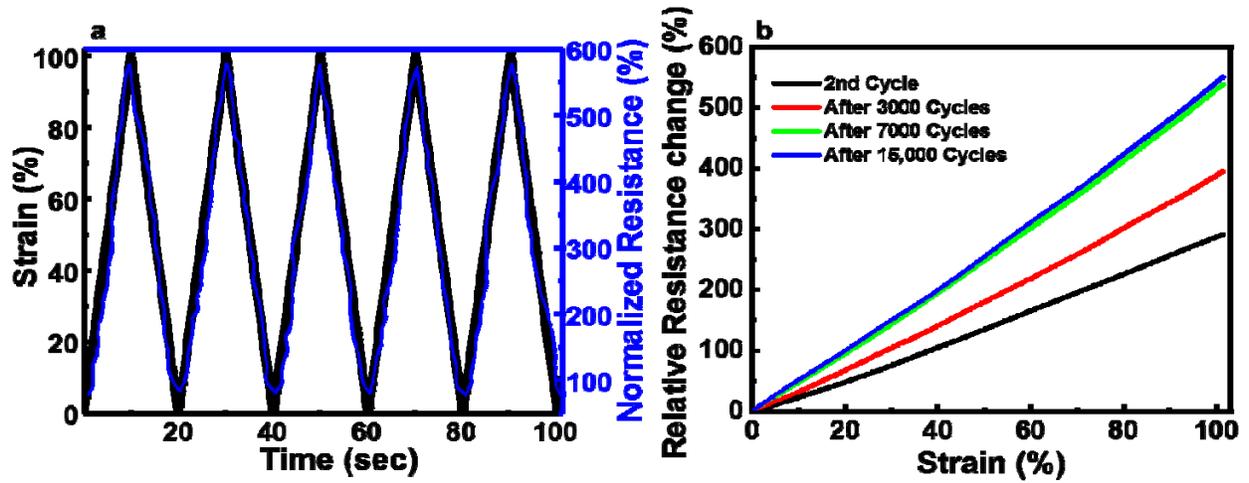

**Figure 5**

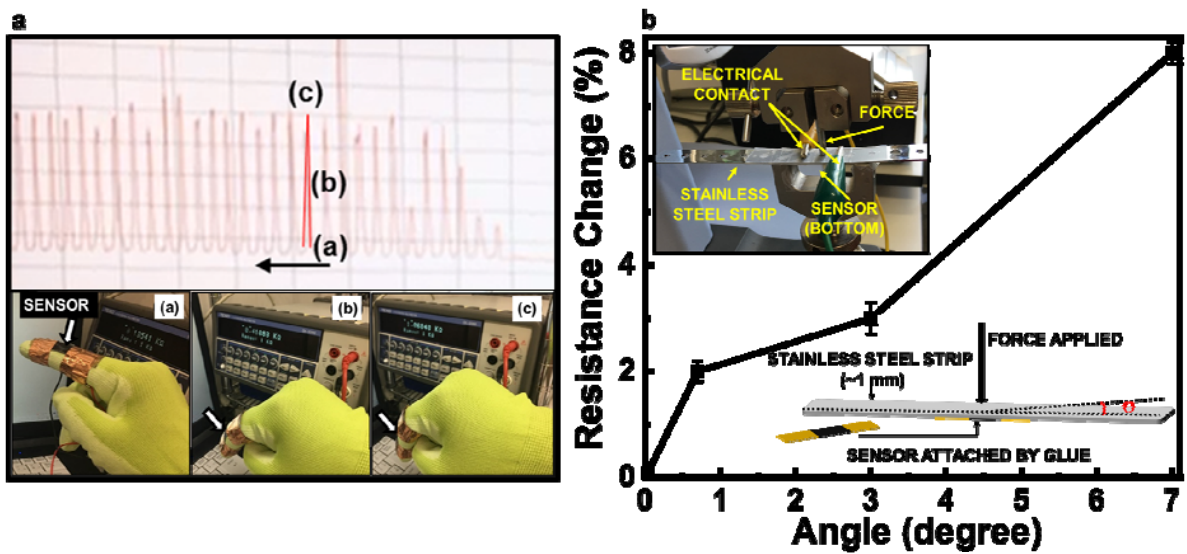

**Figure 6**